\documentclass[letterpaper]{article} 
\usepackage{aaai25}  
\usepackage{times}  
\usepackage{helvet}  
\usepackage{courier}  
\usepackage[hyphens]{url}  
\usepackage{graphicx} 
\usepackage{natbib}  
\usepackage{caption} 
\usepackage{algorithm}
\usepackage{algorithmic}

\usepackage[utf8]{inputenc} 
\usepackage[T1]{fontenc}    
\usepackage{url}            
\usepackage{booktabs}       
\usepackage{amsfonts}       
\usepackage{microtype}      
\usepackage{xcolor}         
\usepackage[utf8]{inputenc}
\usepackage{graphicx}
\usepackage{subcaption}
\usepackage{amsmath}
\usepackage{amsthm}
\usepackage{tcolorbox}
\usepackage{tabularx}
\usepackage{svg}
\usepackage{lipsum} 
\usepackage{xspace}
\usepackage{colortbl}
\definecolor{Gray}{gray}{0.9}

\newtheorem{definition}{Definition}

\newtheorem{theorem}{Theorem}
\newtheorem{lemma}{Lemma}

\newtheorem{property}{Property}
\newtheorem{assumption}{Assumption}
\newtheorem{remark}{Remark}

\newcommand{\SysName}{LSP-Offload\xspace} 
\newcommand{\GPUmodel}{{\cal M}}

\usepackage{newfloat}
\usepackage{listings}
\DeclareCaptionStyle{ruled}{labelfont=normalfont,labelsep=colon,strut=off} 
\floatstyle{ruled}
\newfloat{listing}{tb}{lst}{}
\floatname{listing}{Listing}
\title{Practical Offloading for Fine-Tuning LLM on Commodity GPU\\ via Learned Sparse Projectors}
\author{
    Siyuan Chen\textsuperscript{\rm 1}, 
    Zhuofeng Wang\textsuperscript{\rm 2},
    Zelong Guan\textsuperscript{\rm 1},
    Yudong Liu\textsuperscript{\rm 1},
    Phillip B. Gibbons\textsuperscript{\rm 1}
}
\affiliations{
    \textsuperscript{\rm 1} Carnegie Mellon University,
    \textsuperscript{\rm 2} Peking University \\
    siyuanc3@andrew.cmu.edu,
    wangzf2003@stu.pku.edu.cn,
    zelongg@andrew.cmu.edu,
    yudongltech@gmail.com,
    gibbons@cs.cmu.edu

}

\usepackage{bibentry}

\begin{document}

\maketitle

\begin{abstract}
Fine-tuning large language models (LLMs) requires significant memory, often exceeding the capacity of a single GPU. A common solution to this memory challenge is offloading compute and data from the GPU to the CPU. However, this approach is hampered by the limited bandwidth of commodity hardware, which constrains communication between the CPU and GPU, and by slower matrix multiplications on the CPU.

In this paper, we present an offloading framework, \textit{\SysName}, that enables near-native speed LLM fine-tuning on commodity hardware through \textit{learned sparse projectors}. Our data-driven approach involves learning efficient sparse compressors that minimize communication with minimal precision loss. Additionally, we introduce a novel layer-wise communication schedule to maximize parallelism between communication and computation. As a result, our framework can fine-tune a 1.3 billion parameter model on a 4GB laptop GPU and a 6.7 billion parameter model on a 24GB NVIDIA RTX 4090 GPU.
Compared to state-of-the-art offloading frameworks, our approach reduces end-to-end fine-tuning time by 33.1\%-62.5\% when converging to the same accuracy. We open source our framework at \url{https://github.com/gulang2019/LSP-Offload}.
\end{abstract}

\section{Introduction}

Recent years have highlighted the remarkable success of billion scale LLMs. Hand-to-hand with task performance improvements are the ever-growing model sizes and the strong demand for powerful computing resources that are available only in high-end clusters. Fortunately, fine-tuning provides everyday ML practitioners the accessibility to LLMs by allowing them to adapt a pre-trained model to downstream tasks using less onerous computational effort. However, fine-tuning's memory and compute demands are still daunting. For example, under a default fine-tuning configuration that uses the fp16 data type with the Adam optimizer~\cite{kingma2014adam}, the memory footprint is 8 $\times$ \#Parameters bytes, which means top-notch commodity workstation GPUs (e.g., NVIDIA 4090 GPU and AMD 7900XTX with 24GB memory each) are able to hold only smaller LLMs (3B parameters). With commodity laptop GPUs (e.g., NVIDIA A1000 with 4GB memory), even 0.77B parameter LLMs do not fit.

A variety of techniques have been proposed to reduce the memory demand during fine-tuning. A typical solution from system researchers is to offload part of the compute and memory from GPU to CPU, leveraging the fact that commodity laptop CPUs typically have 4x the memory of laptop GPUs and commodity workstation CPUs can provide 4TBs of memory (per socket). Although offloading is able to scale the trainable model size, large batch sizes are essential to remain efficient despite the limited PCIe bandwidth between CPU and GPU~\cite{zero-inifity}. In fact, we show that training with offloading is inherently bounded by either the CPU-GPU communication or the compute on CPU, especially on commodity hardware where the limited GPU memory dictates small batch sizes. Therefore, offloading itself can hardly save us from the scaling challenge.

Meanwhile, another promising method from ML researchers for memory-reduction is parameter-efficient fine-tuning (PEFT). The key idea of PEFT is to limit the trainable parameters to a carefully designed subspace (e.g., a low rank subspace~\cite{hu2021lora,zhao2024galore} or only part of the model~\cite{guo2020parameter}), so the GPU can train the model without offloading as long as it can hold the parameters and minimal optimizer states for the trainable parameters. However, though more memory-efficient, PEFT methods can suffer from slow convergence or sub-optimal training results due to their overly constrained space for parameter updates.

\begin{figure}
    \centering
    \includegraphics[width=\linewidth]{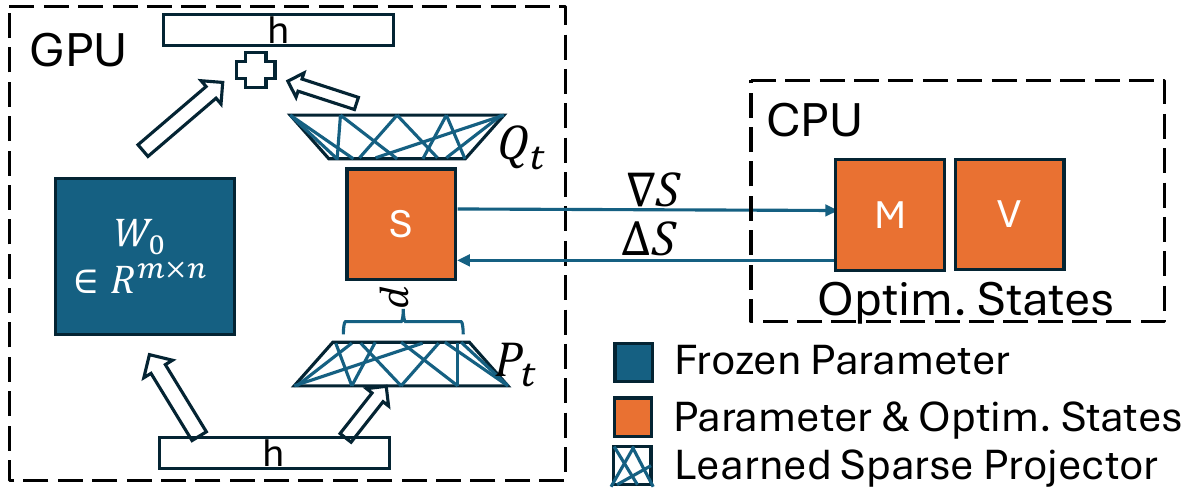}
    \caption{LSP-Offload}
    \vspace{-0.2in}
    \label{fig:lsp-offload}
\end{figure}

In this paper, we show how to mitigate the memory challenge by combining both approaches. We present \SysName (Fig.~\ref{fig:lsp-offload}), a novel fine-tuning framework that (i) mitigates bottlenecks in offloading approaches by a new approach to refactor the offloading process and (ii) trains efficiently by a new approach to constrain the optimization space.

Specifically, to alleviate the compute pressure on the CPU as well as the communication overhead back-and-forth between CPU and GPU, we constrain the updates to happen on a periodically-changing subspace ($S$ in Fig.~\ref{fig:lsp-offload}). Because the updates from different subspaces are projected back and accumulate together in the original space, the model is able to update in the full-rank optimization space. State-of-the-art (SOTA) approaches~\cite{hu2021lora,zhao2024galore} for constraining the parameter update space suffer from linear memory and compute complexity that limits them from optimizing in large subspaces. 
We solve this problem by the introduction of \textit{$(d,r)$-sparse projectors} ($P_t$ and $Q_t$ in Fig.~\ref{fig:lsp-offload}), sparse embedding matrices that represent a subspace but whose memory consumption is independent of the subspace's size. In this way, given the same memory budget as PEFT, we are able to optimize in an arbitrary-size subspace. To further improve the compression quality of the subspace, we adopt a data-driven approach similar to \cite{liu2020framework} that adapts the subspace to the gradient matrices, which is empirically proven necessary for fast convergence. 

Moreover, at the system level, we demonstrate that the SOTA offloading framework \textit{Zero-Offload}~\cite{zero} suffers from limited parallelism between communication and compute when running on commodity hardware. This is due to the limited GPU memory relative to the model size, which implies that only small batch sizes can be used during training. We improve Zero's schedule by performing fine-grained communication on the granularity of layers and communicating components of the gradient ahead of time. The new schedule enables us to explore the full parallelism between CPU compute, GPU compute, CPU-to-GPU communication, and GPU-to-CPU communication.



In summary, our paper makes the following contributions: 
\begin{itemize}
    \item We analyze LLM training on commodity hardware (both laptop and workstation) to show that current offloading workflows are fundamentally bounded by either the communication or the CPU's compute.
    \item We design \SysName to enable near-native speed fine-tuning on commodity hardware. The system is built on the key idea of \textit{learned sparse projectors}, which enables fine-tuning on high-dimensional subspaces with constant memory and compute overhead. We open source our framework at \url{https://github.com/gulang2019/LSP-Offload}.
    \item We verify that \SysName converges to the same accuracy as native training on the GLUE dataset. On the instruction-tuning task, \SysName reduces end-to-end fine-tuning time by 33.1\% to 62.5\% over SOTA offloading, when converging to the same accuracy. Moreover, \SysName improves accuracy by 27.8\% to 30\% over SOTA PEFT approaches on the Alpaca and Humaneval datasets.
\end{itemize}
\section{Background and Related Work}




\paragraph{Memory breakdown for training large language models.} Training a deep learning model requires memory for parameters, activations, and optimizer states. Activations include  intermediate results used in backward propagation. The optimizer states are used by the optimizer to update the parameters. 
Of the three, the memory for parameters ($M_{param}$) and optimizer states ($M_{opt}$) consume most of the memory. When trained with the Adam optimizer and half precision, $M_{param} + M_{opt} \approx 8 \times \#\text{Parameters}$ bytes, which easily exceeds the single GPU's memory for billion-scale models. 

\begin{table*}
\caption{Configurations and timings for training/fine-tuning the llama-7B Model (using fp16) on commodity workstation hardware---the Nvidia RTX 4090 GPU and AMD Ryzen Threadripper 3970X CPU. For UPD, we measure the fused Adam kernel with thread-level parallelism and SIMD optimizations. Bandwidth is the PCIe bandwidth with a pinned memory buffer.}
\label{tab:llama7b}
\vspace{-0.05in}
\centering
\begin{tabular}{llllll}
\hline
\rowcolor{Gray}
Parameters & Optimizer State & Activations & CPU-GPU Bandwidth & \#Layers & GPU Memory \\ \hline
14GB        & 42GB             & 8GB         & 10--20GB/s            & 32       & 24GB\\
\hline\hline
\rowcolor{Gray}
FWD on CPU & BWD on CPU & UPD on CPU & FWD on GPU & BWD on GPU & UPD on GPU\\\hline
1.61s/layer & 3.30s/layer & 0.06s/layer & 1.7ms/layer & 3.5ms/layer & 1ms/layer \\
\hline
\end{tabular}
\end{table*}

\paragraph{Memory offloading.} These techniques~\cite{g10,swapadvisor,zero,zero-inifity,zero-offload} enable training the full model with inadequate GPU memory by utilizing non-GPU memory such as CPU memory or SSDs. Among these, Zero series are the SOTA approaches for fine-tuning large models. Zero-Offload~\cite{zero-offload} offloads the optimizer states and the update step onto the CPU. Compared to other approaches that offload only the memory to CPU and do all computations on GPU, Zero-Offload achieves the optimal communication volume for full parameter training. Nevertheless, we found that Zero's training is severely bottlenecked by the communication (see Fig.~\ref{fig:profile}). Our work is built on top of the Zero series offloading schedule to make it practical for single GPU training with minimal communication overhead.

\paragraph{Parameter-efficient fine-tuning.} PEFT enables pre-trained models to rapidly adapt to downstream tasks with minimal extra memory required. LoRA~\cite{hu2021lora} is among the most popular PEFT techniques by constraining the optimization onto a decomposed low-rank subspace. However, recent works~\cite{lialin2023relora,valipour2022dylora} found LoRA is sensitive to hyperparameter tuning and can struggle with tasks requiring significant change to the base model. To break the low-dimensional constraint of LoRA, GaLore~\cite{zhao2024galore} recently explores a similar idea to ours that periodically changes the subspace computed by singular-value-decomposition (SVD). However, both LoRA and GaLore have the limitation that their algorithms require extra memory and compute linear with the subspace's size (rank), which inherently prevent them from tuning on a higher dimensional subspace.
Our work mitigates this problem via novel subspace projectors whose compute and memory demands are independent of the subspace size, enabling us to achieve better model accuracy by tuning in a larger subspace.

\paragraph{Other methods for memory-efficient training.} Various approaches such as quantization~\cite{dettmers2024qlora} and gradient checkpointing~\cite{chen2016training} have been proposed to reduce the memory demand for training/fine-tuning LLMs. The quantization approach uses data types with fewer bits for training, and is fully compatible with our techniques (we use fp16 in our evaluations). Meanwhile, the gradient checkpointing technique trades computation for memory by recomputing activations during the backward pass. We include this technique in our implementation. 

\section{Motivation}
\label{sec:motivation}
\subsection{Numerical Analysis for Fine-tuning on a GPU}

We motivate our work by an analysis on the fundamental limits of vanilla offloading on a single commodity GPU. We use the example setting of fine-tuning a llama-7B model on a Nvidia RTX 4090 GPU (a commodity workstation GPU), which provides only $24/(14+42+8) = 37.5\%$ of required memory (Tab.~\ref{tab:llama7b}).%
\footnote{A similar analysis, with the same general conclusions, can be done for the GPT2-1.3B model on a commodity laptop GPU, based on Table~\ref{tab:gpt2-1b} in the appendix.}

Current offloading techniques can be categorized into two classes: (i) those that offload only memory to the CPU, and (ii) those that offload both memory and compute to the CPU. The first type is represented by \cite{swapadvisor,g10}, which perform all compute on the GPU while swapping in and out memory on the fly. An example of this type of schedule is shown in Fig.~\ref{fig:pipeline}.c. However, this type of offloading schedule is inherently bounded by the communication under the following observation. For our setting, we need 5.33s communication per iteration, which adds 3.2x overhead compared to the GPU compute even if compute and communication are fully overlapped.

\textbf{Observation.} \textit{Training a model demanding $M_{tot}$ memory on a GPU with only $M_{gpu}$ memory, such that the GPU performs all the computation, requires $\geq M_{tot} - M_{gpu}$ of communication per iteration.},

The second type of offloading schedule divides the workload between the CPU and GPU.  Because of the CPU's limited computing power, only parameter update step (UPD) is suitable to run on the CPU. For example, assigning the FWD+BWD pass of just one layer to the CPU directly adds 4.9s overhead, which is already 3.21x the GPU compute. Moreover, offloading UPD to the CPU\footnote{More specifically, the computation of $\Delta W$ to the CPU---applying these deltas to the model parameters remains on the GPU.} means that the 42GB optimizer state can reside on the CPU, enabling larger models like llama-7B to fit in the GPU memory.




Offloading UPD to the CPU was first realized in Zero-Offload~\cite{zero-offload}, whose schedule is displayed in Fig.~\ref{fig:pipeline}.a (Alg.~\ref{alg:zero} in the Appendix).
In their schedule, $M_{param}$ communication happens every iteration (gradients to CPU, deltas to GPU), which brings the communication overhead to 0.93s.
When there is no overlap between CPU compute and GPU compute (Fig.~\ref{fig:pipeline}.a), the training slowdown is 2.11x.
Moreover, the CPU compute can become the bottleneck for Zero's schedule. For the example setting, UPD on the CPU takes 1.92s per iteration, slowing down training by 2.14x. 

This analysis shows that \textbf{training with offloading is computationally inefficient on modern commodity hardware due to fundamental bottlenecks in communication and/or CPU compute.} This motivates us to design a lossy (PEFT) algorithm for reduced overheads when offloading.

\subsection{Case Study on Zero's Schedule}

\begin{figure}
    \centering
    \vspace{-0.1in}
    \includegraphics[width=0.4\textwidth]{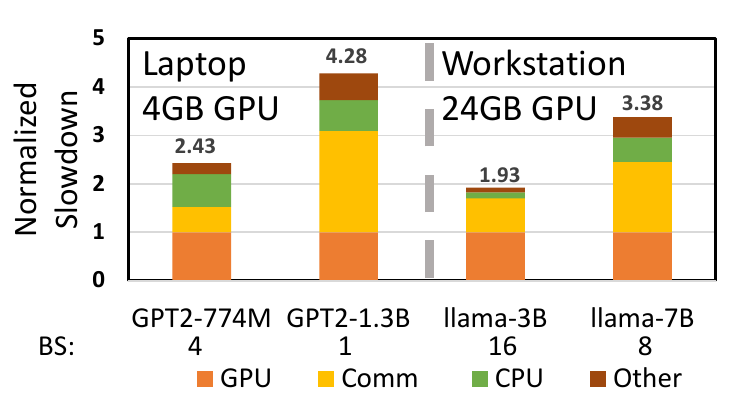}
  \vspace{-0.1in}
  \caption{Normalized slowdown of Zero's schedule on laptop and workstation GPUs. The breakdown for communication (Comm) depicts the additional slowdown due to communication that is \textbf{not} overlapped with GPU compute. Similarly, the CPU compute and Other are additional non-overlapped overheads. The experiments are done using precision fp16, the largest batch sizes (BS) that fit, and gradient checkpointing.}
    \label{fig:profile}
    \vspace{-0.22in}
\end{figure}
Moreover, prior offload schedules are suboptimal. We profile Zero-Offload's schedule for a more comprehensive view of its performance. We study two settings for profiling: (i) training a GPT2 model on a 4GB laptop GPU, and (ii) training a llama model on a 24GB workstation GPU. The slowdown normalized by the GPU compute time is shown in Fig.~\ref{fig:profile}. 
Under both configurations, Zero's schedule slows training by 1.93x to 4.28x, for the following two reasons.

\begin{figure*}[t]
    \centering
    \includegraphics[width=\textwidth]{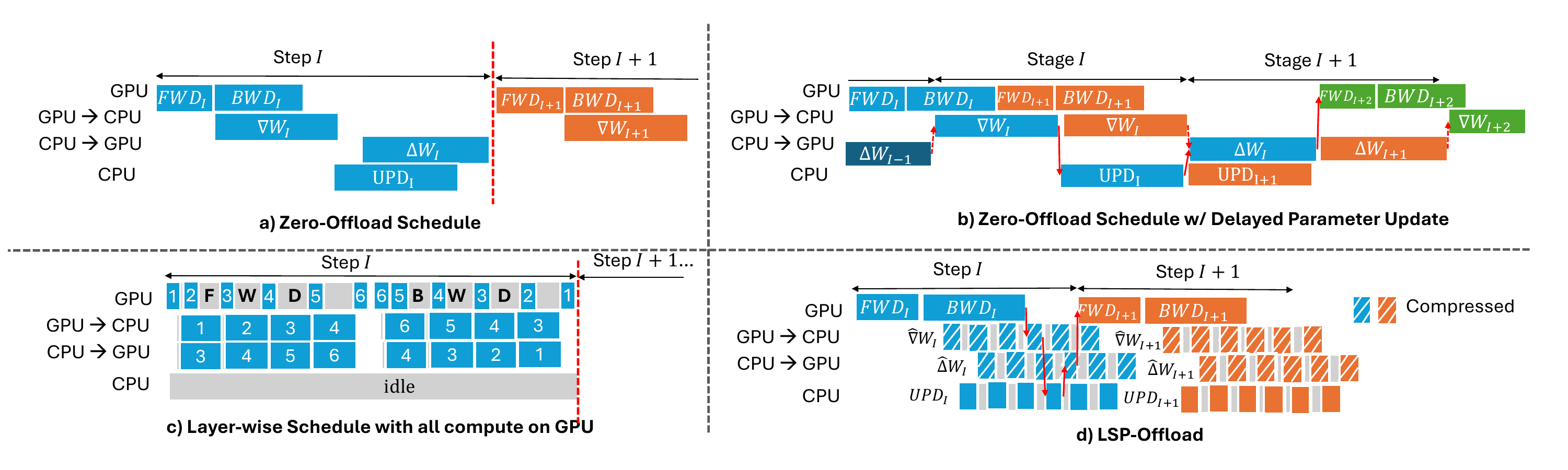}
    \vspace{-0.3in}
    \caption{Comparison between current offloading pipelines and \SysName's overlapped pipeline.}
    \vspace{-0.20in}
    \label{fig:pipeline}
\end{figure*}

\paragraph{Communication and CPU compute overhead.} The primary source of overhead comes from the unavoidable high communication volume and slow CPU compute as demonstrated in our previous analysis. Shown in Fig.~\ref{fig:profile}, although Zero is able to overlap part of the GPU/CPU compute with communication, the non-overlapped communication brings 0.61x to 2.09x added slowdown compared to the GPU compute time. 
For both the laptop and workstation GPUs, the situation is worse for the larger model because the maximum available batch size decreases. 
When training a 1.3B model on a 4GB GPU, the non-overlapped communication and CPU compute are 2.09x, 0.63x the GPU compute, respectively. 

\paragraph{Limited parallelism between CPU and GPU, communication and compute.} The second source of overhead comes from Zero's limited parallelism between compute and communication. 
Fig.~\ref{fig:pipeline}.a shows Zero's standard training pipeline, which is suboptimal for two reasons: (i) FWD and BWD on the GPU are not overlapped with the CPU's compute. This results in significant slowdown when the CPU compute is around the same scale as the GPU compute. (ii) There is no overlap between the GPU-to-CPU communication and the CPU-to-GPU communication, which implies that the full duplex PCIe channel is at least 50\% underutilized. 
As a result, the per-iteration time for Zero's schedule is
\begin{multline}
T_{Zero}^{iter} = T_{FWD} + \max\{T_{BWD}, T_{Comm}^{\text{GPU-to-CPU}}\} + \\
\max\{T_{UPD}, T_{Comm}^{\text{CPU-to-GPU}}\}. \label{eqn:zero-nodelay}
\end{multline}

To mitigate the first issue, Zero proposed delayed parameter updates (Fig.~\ref{fig:pipeline}.b), which use stale parameter values to calculate current gradients, allowing the CPU to perform the previous step's update at the same time the GPU performs the current step's forward and backward passes. 
Although increasing throughput, this method can affect the accuracy of training. Also, in order not to incur additional memory for buffering communication, the CPU-to-GPU communication and GPU-to-CPU communication cannot be parallelized. 

These limitations inspire our design of a layer-wise scheduling strategy that maximizes parallelism between computation and communication. Unlike prior works that focus on parameter pulling or collective communication~\cite{wang2019scalable} in distributed training~\cite{lee2017parallel}, our approach applies layer-wise overlapping to offloading, achieving optimal parallelization across CPU and GPU computations and their communications.
\section{\SysName's Approach}
In this section, we present \SysName, a practical offloading framework for fine-tuning high-quality models efficiently under memory-constrained settings. We will introduce our training algorithm for mitigating the compute and communication overhead, and then illustrate our new schedule design for maximized parallelism in the offloading's schedule.


\subsection{Efficient and High-quality Offloading via Learned Sparse Projectors}
As discussed before, on commodity hardware, the large optimization space combined with limited communication bandwidth causes offloading with a standard training algorithm to result in significant communication and compute overheads. To mitigate this problem, our key insight is to assist the offloading algorithm by using PEFT to configure the size of the optimization subspace, but to do so using novel techniques that avoid the pitfalls of prior PEFT. 



Fig.~\ref{fig:lsp-offload} illustrates our approach. Following previous works~\cite{hu2021lora, zhao2024galore}, we focus on matrix multiplication operations. Similar to LoRA and GaLore, we freeze the pre-trained weight matrix and optimize on a decomposed subspace. 
However, the rank of LoRA's and GaLore's optimization space is linearly growing with the extra GPU memory needed to store the projectors and the optimization states, preventing them from optimizing in a sufficiently large subspace. 
E.g., as shown in \cite{zhao2024galore}, fine-tuning a 1B model with a hidden size of 2048 on a rank-512 subspace in half precision requires 4.38GB for LoRA and 6.17GB for GaLore, adding 119\% and 208\% GPU memory overhead compared to storing only the pre-trained model.  

\begin{table}[t]
\caption{Comparison between different fine-tuning approaches, where $n, d, r$ are tensor dimensions satisfying $n \gg d \gg r$. $W\in R^{m\times n}$ is the frozen pre-trained weight matrix. $\beta \geq 1$ is the scale factor for storing the optimizer state ($\beta=3$ for Adam), $\tau$ is the number of updates on the subspace, and $\gamma_1,\gamma_2 \in (0,1]$ are scaling factors that adjust the rank based on how the individual subspaces interact when added together. \SysName both reduces GPU memory and increases the optimization space rank. }
\vspace{-0.05in}
\label{tab:method comparisons}
\centering
\resizebox{\linewidth}{!}{
\begin{tabular}{@{}l|l|l|l@{}}
\toprule
                          & \textbf{LoRA}                      & \textbf{GaLore}                            & \textbf{\SysName}                 \\ \midrule
Weight Matrix             & $W + AB^T$                  & $W + A_t B_t^T$                   & $W + P_t^T S_t Q_t$         \\ 
Trainable Parameters      & $A, B \in R^{m\times r, n \times r}$ & $B_t \in R^{n \times r}$        & $S_t \in R^{d\times d}$     \\ \midrule
GPU Memory                & $mn+\beta (m+n)r$               & $mn+(m + \beta n) r$                  & \textbf{$mn+(m+n)r$}                       \\ 
Rank(Optim. Space)      & $r$                       & $\gamma_1 r \tau$                 & \textbf{$\gamma_2 d \tau$}           \\
\bottomrule
\end{tabular}
}   
\end{table}

To overcome this limitation, we made the key innovation to design the projector as sparse matrices, decoupling the dependence between the GPU memory overhead and the rank of the optimization space. Specifically, we use $(d,r)$-sparse projectors as the template projector (see the properties of this projector in the appendix).

\begin{definition}[$(d,r)$-Sparse Projector]
We define the projection bases $P\in \mathbb{R}^{m\times d}, Q \in \mathbb{R}^{n\times d}$ as $(d,r)$-sparse projectors if both $P, Q$ have $r$ nonzero values per row. 
\end{definition} 

\begin{figure}
    \centering
    \includegraphics[width=1.0\linewidth]{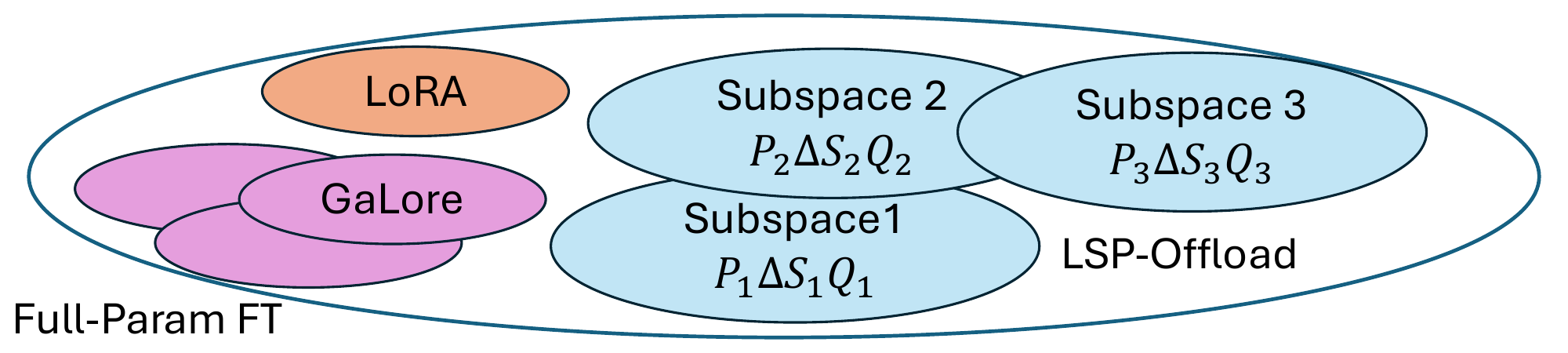}
    \caption{Visualization on Optimization Space.}
    \vspace{-0.1in}
    \label{fig:opt space}
\end{figure}

As shown in Fig.~\ref{fig:lsp-offload}, by using $(d,r)$-sparse projectors to replace the dense projectors, we project the weights on a $d\times d$ dimensional subspace. Meanwhile, the sparsity allows us to store only the $O((m+n)r)$ non-zero values of the projectors on the GPU. This brings \SysName two benefits: 
\begin{itemize}
    \item \SysName is capable of optimizing in a larger subspace while using less GPU memory than SOTA PEFT. For our example setting, \SysName only uses 2.015GB GPU memory when using $r=4$.
    \item \SysName's optimization space scales linearly with the parameter size. \SysName optimizes in subspaces of size $O(d^2)$, with $d$ set to $n/2$ to hide communication overhead. This results in a scaling of $O(n^2)$ as the model size grows, outperforming LoRA and Galore's $O(n\times r)$ scaling, especially when $n >> r$ for large models.
\end{itemize}


In all, the optimization space for a matrix multiplication operation with pre-trained matrix $W_0 \in R^{m\times n}$ constrains as 
\begin{align}
\Delta W &= P_1 S_1 Q_1^T + P_2 S_2 Q_2^T + ... + P_\tau S_\tau Q_\tau^T,
\end{align}
where $P_t, Q_t \in R^{n\times d}$ are periodically updated $(d,r)$-sparse projectors, and $S_t \in R^{d\times d}$ is a dense trainable matrix. Visualized in Fig.~\ref{fig:opt space}, \SysName optimizes in a larger subspace
than LoRA and GaLore while introducing same GPU memory overhead, underscoring \SysName's efficiency.


\begin{algorithm}[t]
\small
\caption{\SysName's fine-tuning with learned sparse projectors [simplified version without layer-wise scheduling]}\label{alg:SGESchedule}
\begin{algorithmic}[1]
\STATE {\textbf{HyperParam:}} $s$: subspace size. $d, r$: $d, r$-sparse projectors. $\gamma \in \mathbb{R}^+$. $CheckFreq, \alpha$: check frequency and threshold for updating projectors. 

\STATE \textbf{Function}\ \textsc{MaybeUpdate}($\nabla_W$: the gradient, $P_{prev}$, $Q_{prev}$: previous projectors, $M$, $V$: optimizer state)\label{line:maybeupdate}
\STATE \hspace{1em}\textbf{if} {$\|\textbf{b}^{P,Q}(\nabla_W)\|_F / \|\nabla_W\|_F \le \alpha $} \textbf{then} \label{line:optional update}
\STATE \hspace{2em} \textbf{Return} $P_{prev},Q_{prev}$
\STATE \hspace{1em} $P, Q \gets Initialize(d,r)$
\STATE \hspace{1em} Minimize $loss := \|\textbf{b}^{P,Q}(\nabla_W)\|_F + \beta \cdot (\|P\|_F^2 + \|Q\|_F)$ until $\|\textbf{b}^{P,Q}(\nabla_W)\|_F / \|\nabla_W\|_F \le \alpha$ or Timeout. \label{line：minimize subspace}
\STATE \COMMENT{Project previous M and V tensors to new subspace:}
\STATE \hspace{1em} $M \in \mathbb{R}^{s\times s} \gets P^TP_{prev} M Q_{prev}^TQ$
\STATE \hspace{1em} $V \in \mathbb{R}^{s\times s} \gets (P^TP_{prev})^2 V (Q_{prev}^TQ)^2$
\STATE \hspace{1em} \textbf{Return} $P, Q$

\STATE \textbf{Function} \textsc{Main}($\GPUmodel$: Model, $\mathcal{D}$: Dataset, $W\in \mathbb{R}^{m\times n}$: Weights, $M, V \in \mathbb{R}^{s\times s}:$ 1st, 2nd order optimizer state)
\FOR{$t \gets 0$ to $\tau - 1$}
     \STATE Sample $x\sim \mathcal{D}$
    \STATE $\nabla_W \gets forwardBackward(\GPUmodel, x)$\COMMENT{FWD+BWD on GPU}
    \STATE $grad \gets SendToCPU(P^T\nabla_W Q)$\COMMENT{Compress on GPU and gradient offload} \label{line:project}
    \STATE $\Delta_{W} \gets SendToGPU(Update(grad))$\COMMENT{UPD on CPU and delta upload} \label{line:update}
    \STATE $W \gets W - \eta_t P\Delta_{W}Q^T$\COMMENT{Decompress, apply deltas on GPU} \label{line:accumulate}
    \IF{$t \bmod CheckFreq = 0$} \label{line:periodical update}
        \STATE $\nabla_W \gets$ gradient on sampled subset $\mathcal{D'} \subset \mathcal{D}$.
        \STATE $P, Q \gets$ \textsc{MaybeUpdate}($\nabla_W$, $P$, $Q$, $M$, $V$)
    \ENDIF \label{line:update end}
\ENDFOR
\end{algorithmic}

\end{algorithm}

\paragraph{Training algorithm.} The above design leads to the \SysName's core training algorithm listed in Alg.~\ref{alg:SGESchedule}. In every iteration, the gradient is projected onto a subspace (line~\ref{line:project}) before transferred to the CPU. The weight delta is then computed on CPU by optimizing on the subspace (line~\ref{line:update}) before transferred back to GPU and projected to the original space (line~\ref{line:accumulate}). 
This way, both communication and compute complexity for offloading is reduced from $O(m\cdot n)$ to $O(d^2)$, which guarantees our algorithm's efficiency.
Moreover, we optionally update the subspace (lines~\ref{line:periodical update}-\ref{line:update end}) by checking its quality.
In the next section, the steps are further pipelined between layers to hide the latency. 
Next, we introduce several techniques to boost the training quality.


%

\paragraph{Learned sparse projectors.} First, we boost the performance of the sparse projectors with a data-driven approach. 
Specifically, we initialize the $(d,r)$-sparse projectors by randomly sampling the $r$ nonzero positions for each row and randomly sampling the nonzero values from $\mathcal{N}(0, 1/\sqrt{r})$. Random sampling ensures an unbiased estimation gradient with good approximation properties, as supported by the JL lemma~\cite{kane2014sparser}. After that, we fit the projectors on the calibration dataset to minimize the following estimation bias on the gradient:
\begin{definition}[estimation bias]\label{def:est bias}
For a $(d,r)$-Sparse Projector $P$, $Q$ and a matrix $\Sigma \in R^{m\times n}$, the estimation bias is
$\textbf{b}^{P,Q}(\Sigma) :=  PP^T\Sigma QQ^T - \Sigma$.
\end{definition}

Denote the forward pass of the matrix multiplication operation as $Wx=(W_0 + P S Q^T)x$. We optimize the following problem for better projectors:    
\begin{align}
\min_{P,Q} \underbrace{\|\textbf{b}^{P,Q}(\nabla_W)\|_F}_{\text{estimation error of gradient}} + \beta \cdot \underbrace{(\|P\|_F + \|Q\|_F)}_{\text{regularization}}
\end{align} 

Compared to GaLore, which uses SVD decomposition as the projection matrix, we empirically find that our data-driven approach has a lower generalization error when using the same amount of extra GPU memory (Fig.~\ref{fig:hyperparameter_testloss}).


\paragraph{Updating the subspace.} Secondly, we avoid the overhead of frequently training the projectors by optionally updating the subspace. Specifically, on a subsampled dataset, only when the gradient estimation bias exceeds a certain threshold $\alpha$ (line~\ref{line:optional update}), do we switch to a new (learned) projector. 

\paragraph{Convergence Analysis of Alg.~\ref{alg:SGESchedule}.} 
For dataset $\mathcal{D}$, weight matrix $W\in R^{m\times n}$, we consider minimizing $f(W) = \Sigma_{x\sim \mathcal{D}} f_x(W) / |\mathcal{D}|$ using Alg.~\ref{alg:SGESchedule} with $CheckFreq=1$. That is, $W_{t+1} = W_t - \eta P_t P_t^T \nabla f_{x_t}(W_t) Q_t Q_t^T, t = 1, 2, ..., T$, where $P_t, Q_t$ are $(d,r)$-sparse projectors.
We derive the convergence theorem based on L-smooth functions, which indicates convexity and smoothness and is widely used in previous studies~\cite{stich2020analysis,garrigos2023handbook}.  
\begin{assumption}[Effectiveness of the subspace]\label{asp:effOfSubspace}
    The relative error on the subspace is kept under $\alpha$ in Alg.~\ref{alg:SGESchedule}.
\end{assumption}

\begin{assumption}[Bounded Bias]\label{asp:boundedBias}
    There exists $\gamma > 0$,  such that for any weight $W$ and $x\sim \mathcal{D}$, $\|\textbf{b}^{P_t,Q_t}(\nabla f_x(W))\| < \gamma, \|\nabla f_x(W)\| < \gamma$.
\end{assumption}

\begin{assumption}[Sparse Bias]\label{asp:sparseBias}
    There exists a constant $0 < c < \frac{1}{\sqrt{2}\alpha}$, such that $\|\textbf{b}^{P_t,Q_t}(\nabla f(W))\|_F < c\|\textbf{b}^{P_t,Q_t}(\nabla f(W))\|_2$ holds for any weight matrices $W$.
\end{assumption}

We show the following convergence rate of our algorithm---please see the appendix for the proof. The key idea is that a small gradient estimation error on the full dataset, which drives convergence, can be inferred from a bounded gradient estimation error on the sub-sampled dataset.

\begin{theorem}\label{the:convergence}
    For any $\beta >0, 0 < \delta < 1$, suppose that f is an L-smooth function, Assumptions~\ref{asp:effOfSubspace},~\ref{asp:boundedBias},~\ref{asp:sparseBias} hold and that we check every iteration in Alg.~\ref{alg:SGESchedule} with the subsampled data set $\cal{D}'$ of size $\mathcal{O}(\frac{8\gamma^2}{3\beta^2}\log{\frac{(m+n)T}{\delta}})$, and stepsize $\eta = \frac{1}{L}$. Denote $F:= \mathbb{E}[f(W_0)] - f^*$. Then with probability $1 - \delta$, 
    $\tau=\mathcal{O}(\frac{1}{\epsilon})\cdot \frac{LF}{(1-2c^2\alpha^2)}$
iterations are sufficient to obtain $\min_{t\in[T]}\mathbb{E}\|\nabla f(W_t)\|^2 = \mathcal{O}(\epsilon + \frac{2c^2\beta^2(1+\alpha)^2}{1 - 2c^2\alpha^2})$.
\end{theorem}

\begin{remark}
     The quality of the subspace ($\alpha$) is critical both for the final accuracy and for the time to convergence.
\end{remark}

\begin{remark}
    The logarithmic sample efficiency in optional update approach indicates low overhead of subsampling ${\cal D'}$.
\end{remark}





\subsection{Layer-wise Schedule for Maximal Parallelism}

At the system level, we propose a new scheduling approach that addresses both issues in Zero's schedule, based on the observation that \textit{ optimization update steps for different layers are independent}. This allows us to overlap GPU computation, CPU-GPU communication in both directions, and parameter updates on the CPU across different layers.
The key idea and its benefits are illustrated in Fig.~\ref{fig:pipeline}.d (Alg.~\ref{alg:LayerwiseSch} in the appendix presents pseudocode).
We split the GPU-to-CPU, CPU update, and CPU-to-GPU communication into small blocks to unlock the parallelism between layers without the accuracy loss of Zero's use of stale parameter values.
We parallelize the CPU's and GPU's compute by executing the deeper layers' update step on CPU while doing the backward pass of shallower layers on GPU.
We also parallelize the double-sided communication by executing deeper layer's upload step while doing the shallower layer's offload step. 
Thus, in our schedule, the critical path of the training is characterized by
\begin{multline}
    T_{LSP}^{iter} = \max\{T_{FWD}+T_{BWD} + T_{Comm}^{layer} + T_{UPD}^{layer}, \\
    T_{Comm}^{\text{GPU to CPU}}, T_{Comm}^{\text{CPU to GPU}}, T_{UPD}\}.
\end{multline}
Compared to Eqn.~\ref{eqn:zero-nodelay},
\SysName reduces the CPU's involvement in the critical path from the entire parameter update step to the update for only one layer, a 32x improvement for the llama-7B model. We show in appendix how to avoid a deeper layer's workload from blocking a shallower layer's computation that executes earlier in the next iteration.

\section{Evaluation}

We first verify the convergence of \SysName on the GLUE dataset and then evaluate the end-to-end training performance on the instruction-tuning task.  Detailed configurations for the experiments are described in the Appendix. 

\paragraph{Accuracy validation of \SysName on GLUE.}
Tab.~\ref{tab:glue} summarizes the accuracy of \SysName for fine-tuning the pre-trained RobertA-base \cite{liu2019roberta} (117M) model on the GLUE~\cite{wang2018glue} dataset, which is a language understanding task set that is widely adopted for evaluating fine-tuning~\cite{hu2021lora,zhao2024galore}. For hyperparameters, we set both the rank of GaLore's projector and the non-zero entries per row in LSP to be 16, so that they use equal GPU memory. The projection space of LSP is set to 512. As both GaLore and LSP need additional profiling time, we make an end-to-end comparison that allows all candidates to train under an hour's time budget. 

As shown in Tab.~\ref{tab:glue}, \SysName outperforms full parameter tuning by 1.9\% accuracy, despite using only 253MB GPU memory vs.~747MB. Furthermore, Fig.~\ref{fig:glue} in the Appendix shows that \SysName converges at the same rate with full parameter tuning. Compared to Galore, \SysName achieves 1.2\% higher average accuracy. We attribute this to \SysName's larger parameter update space (for the same GPU memory), which is 10x for this experiment.

\begin{table}[]
\caption{Accuracy validation of LSP after 1 hour fine-tuning the pre-trained RoBertA-base model on GLUE.}
\vspace{-0.1in}
\resizebox{\linewidth}{!}{
\begin{tabular}{@{}l|llllllll|l@{}}
\toprule
                              & {MNLI} & {SST2}  & {MRPC}  & {CoLA}   & {QNLI}   & {QQP}   & {SST2}  & {STS-B} & {Avg}       \\ \midrule
Full Parameter                & 0.8111        & \textbf{0.934} & 0.866 & 0.55   & 0.904  & 0.808          & \textbf{0.933} & 0.884 & 0.8362625 \\ 
GaLore (Rank=16)            & \textbf{0.83} & 0.92           & 0.88           & 0.567           & 0.881           & \textbf{0.852} & 0.92           & 0.9            & 0.84375            \\ 
LSP ($d$=512, $r$=16) & 0.814         & 0.917          & \textbf{0.911} & \textbf{0.6165} & \textbf{0.9178} & 0.8339         & 0.922          & \textbf{0.91}  & \textbf{0.855275} \\
\bottomrule
\end{tabular}
}
\label{tab:glue}
\vspace{-0.15in}
\end{table}



\begin{table}
\caption{Evaluation accuracy on the Humaneval dataset instruction after fine-tuning Deepseek-Coder-1.3B (top) and Deepseek-Coder-6.7b (bottom) with bfloat16 on the laptop GPU (top) and workstation GPU (bottom).}
\label{tab:coding}
\vspace{-0.1in}
\resizebox{\linewidth}{!}{
\begin{tabular}{@{}l|rr|rrrrrr|r@{}}
\toprule
 & GPU Mem & Time & python & java & cpp & js & ts & php & Avg. \\
\midrule
Zero-Offload & 3.3GB & 120h & \textbf{57.93} & 37.97 & 39.75 & \textbf{52.80} & 47.17 & \textbf{40.99} & 45.5 \\
LoRa (Rank=8) & 3.6GB & 120h & 43.29 & 41.77 & 35.40 & 41.61 & 43.40 & 31.68 & 39.3 \\
GaLore (Rank=256) & 7.9GB  & 120h & 39.63 & 36.08 & 31.68 & 34.78 & 40.88 & 36.02 & 36.4 \\
LSP ($d$=1280, $r$=4) & 3.6GB & 120h & 55.49 & \textbf{42.41} & \textbf{40.99} & 50.31 & \textbf{48.43} & 38.51 & \textbf{45.6} \\
\midrule
Zero-Offload & 16.8GB & 15h & 73.78 & 61.39 & \textbf{64.60} & 66.46 & 64.15 & 58.39 & 64.8  \\
Zero-Offload & 16.8GB & \color{red}{30h} & \textbf{75.00} & \textbf{64.56} & 61.49 & \textbf{70.81} & 65.41 & 62.73 & \textbf{66.7} \\
LSP ($d$=2048, $r$=8) & 17.0GB & 15h & 74.39 & 62.66 & 61.49 & 66.46 & \textbf{67.30} & \textbf{65.84} & 66.4 \\
\bottomrule
\end{tabular}
}
\vspace{-0.2in}
\end{table}


\begin{figure*}[ht]
     \centering
     \begin{subfigure}[t]{0.24\textwidth}
         \centering
         \includegraphics[width=\textwidth]{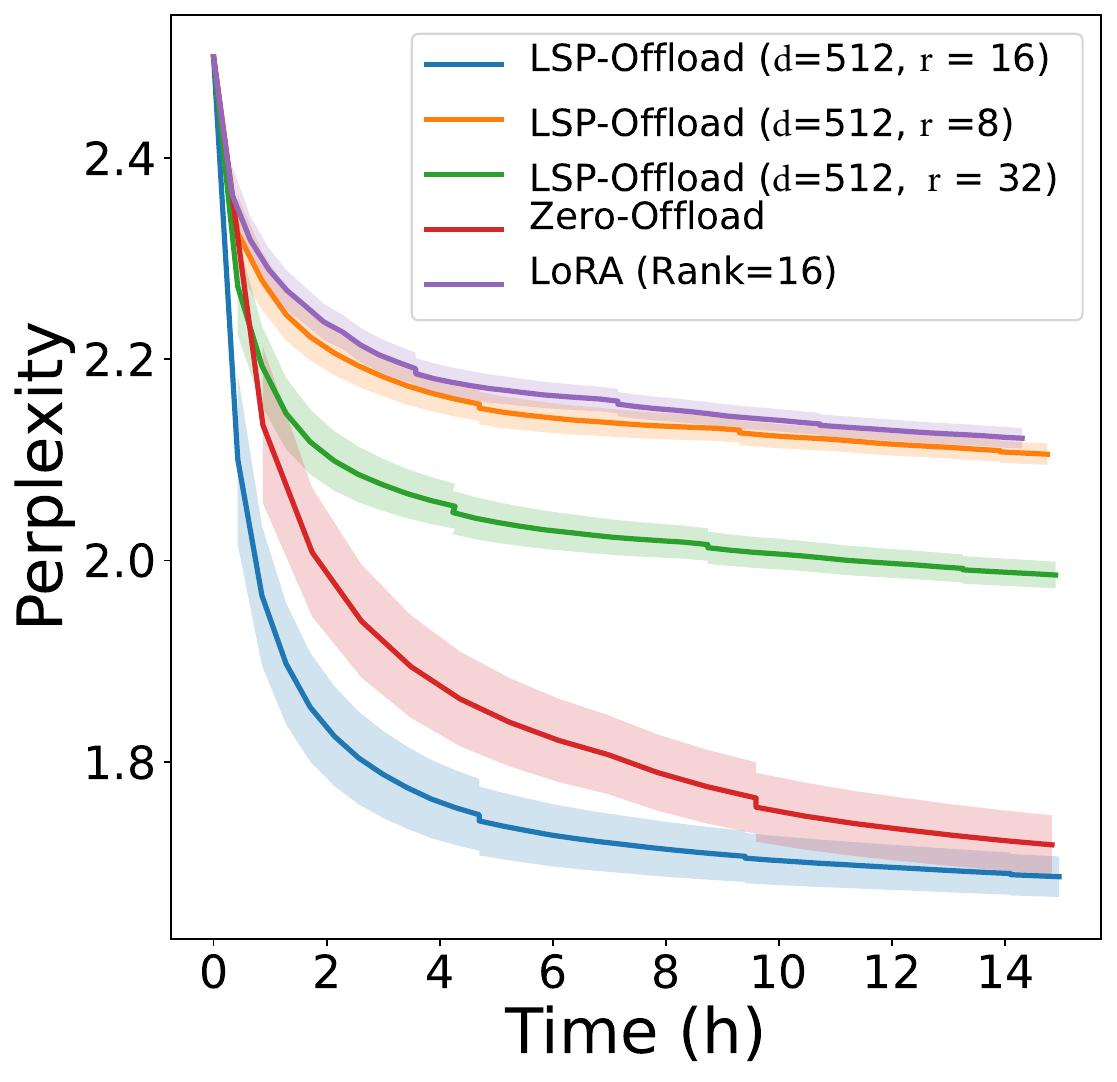}
         \caption{Evaluation perplexity of fine-tuning GPT2-774M w/ the laptop GPU. }
         \label{fig:gpt2}
     \end{subfigure}
     \hfill
     \begin{subfigure}[t]{0.24\textwidth}
         \centering
         \includegraphics[width=\textwidth]{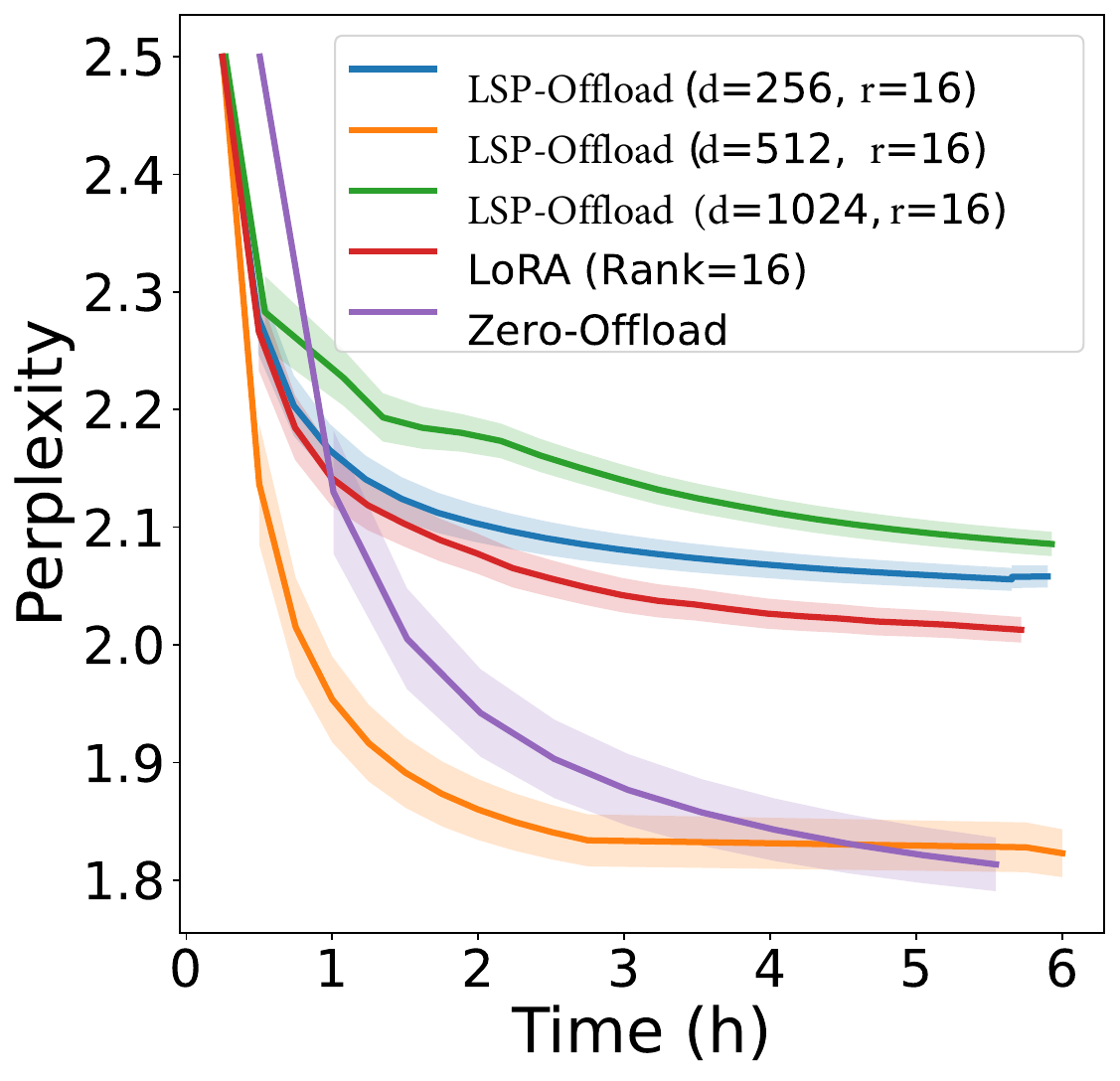}
         \caption{Evaluation perplexity of fine-tuning Llama-3B w/ the workstation GPU. }
         \label{fig:llama3b}
     \end{subfigure}
     \hfill
     \begin{subfigure}[t]{0.245\textwidth}
         \centering
         \includegraphics[width=\textwidth]{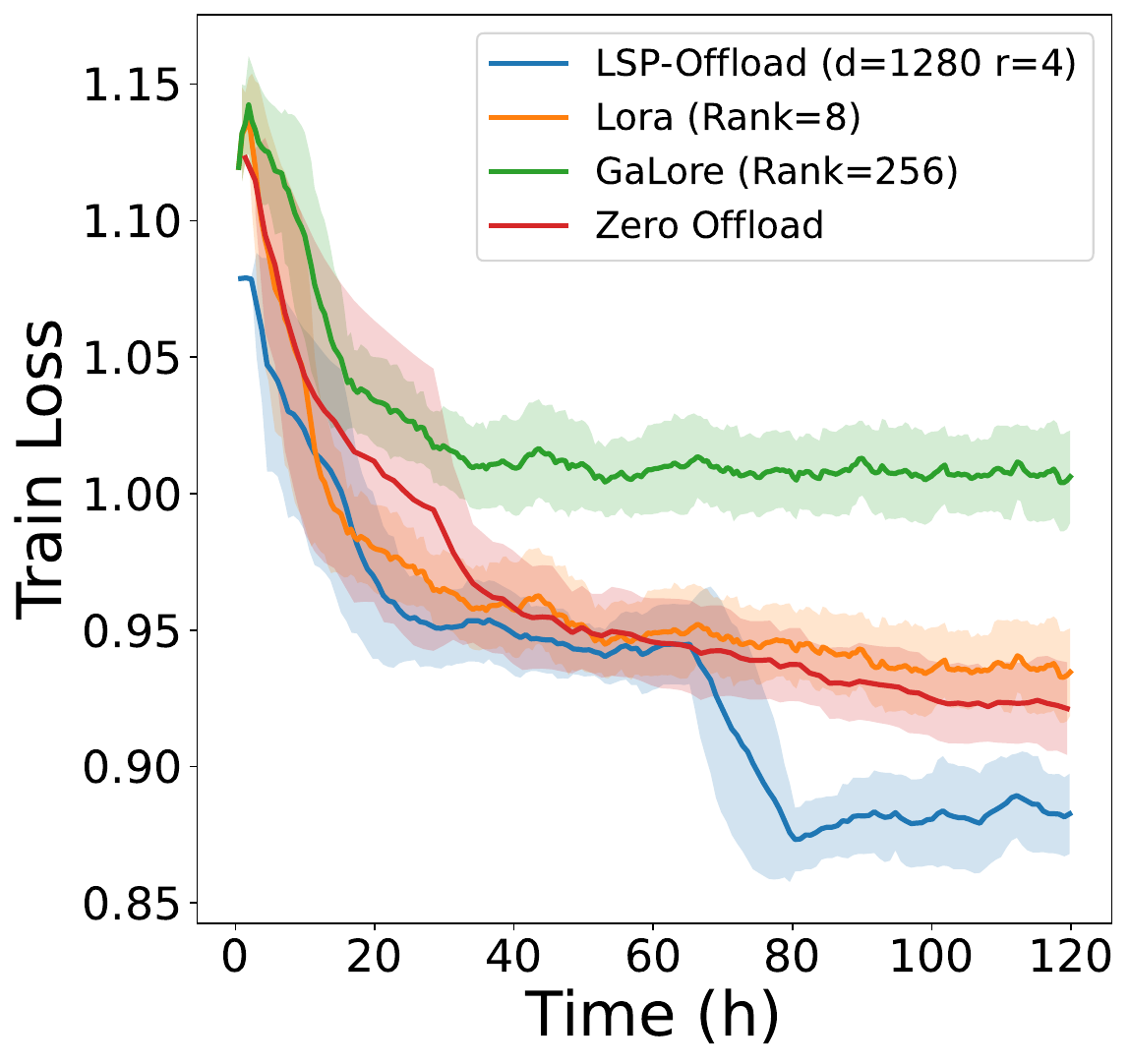}
         \caption{Simulated training loss of fine-tuning Deepseek-Coder-1.3B w/ the  laptop GPU.}
         \label{fig:deepseek-coder}
     \end{subfigure}
     \hfill
     \begin{subfigure}[t]{0.245\textwidth}
         \centering
         \includegraphics[width=\textwidth]{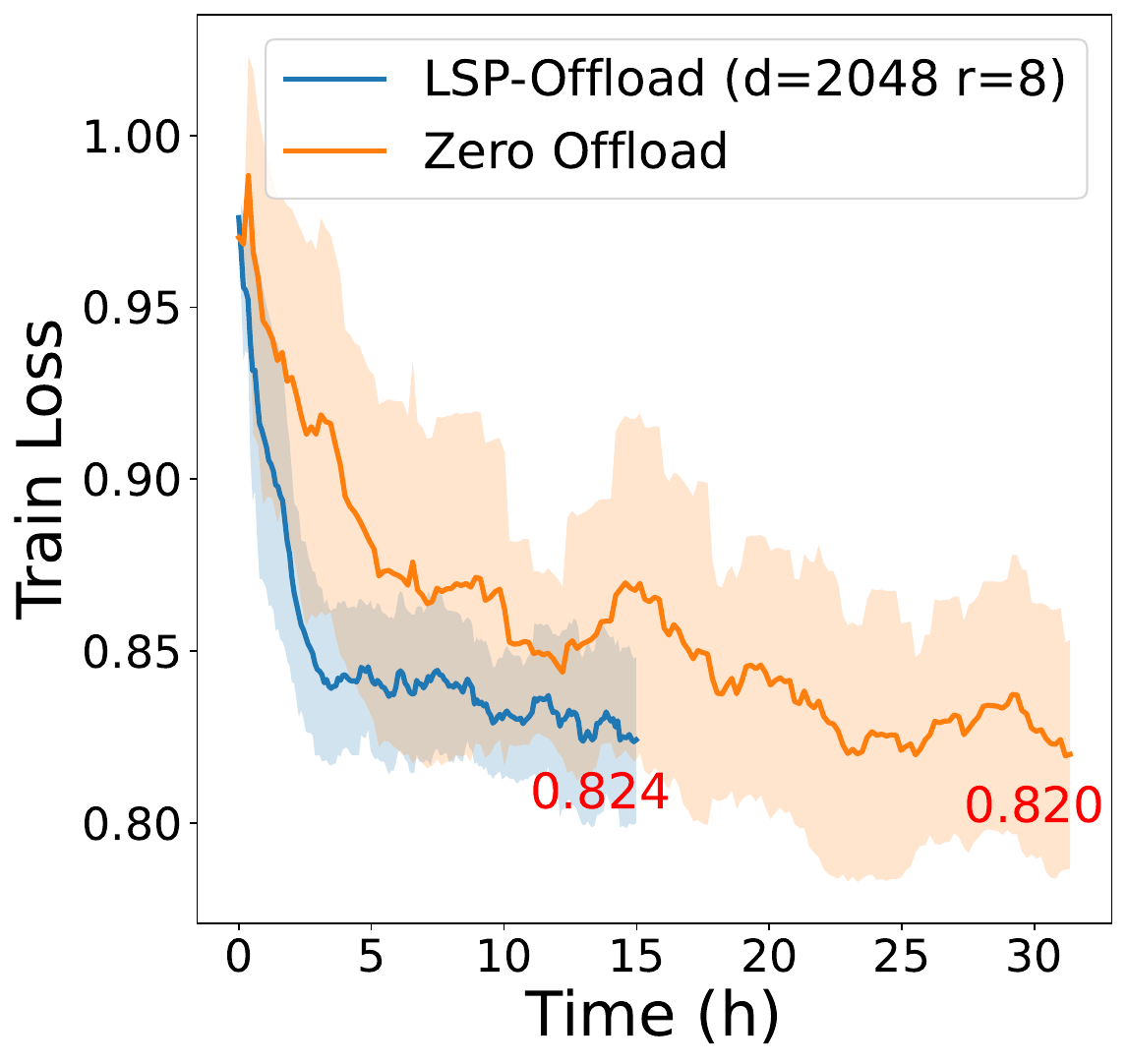}
         \caption{Simulated training loss of fine-tuning Deepseek-Coder-6.7B w/ workstation GPU for one epoch.}
         \label{fig:deepseek-coder-workstation}
     \end{subfigure}
     \vspace{-0.15in}
     \caption{End-to-end evaluation of \SysName. Rolling average is applied. Shades are for deviation.} 
        \label{fig:alpaca}
        \vspace{-0.25in}
\end{figure*}

\paragraph{End-to-end evaluation.}
Next, we evaluate the end-to-end performance of \SysName for instruction-tuning. We perform our evaluation using four settings: (1) fine-tuning the GPT2-774M model on the Alpaca dataset~\cite{alpaca} on a laptop with Nvidia A1000 Laptop GPU (4GB) and Intel Core-i7 12800H CPU (32GB), (2) fine-tuning the Llama-3B model on Alpaca on a workstation with Nvidia RTX 4090 GPU (24 GB) and AMD Ryzen Threadripper 3970X CPU (252GB), and (3,4) fine-tuning the Deepseek-Coder-1.3B model (Deepseek-Coder-6.7B model) on an open-source code instruction dataset generated using WizardCoder's~\cite{luo2023wizardcoder} method
on the laptop GPU (workstation GPU).We choose rank in LoRA, Galore, and $r$ in \SysName such that they use similar amount of memory below the GPU memory capacity. 

\paragraph{Convergence Accuracy}  Shown in Fig.~\ref{fig:deepseek-coder}, \SysName’s final training loss and evaluation accuracy (0.824 / 66.4) closely match those of full parameter tuning (0.820 / 66.7), validating \SysName's minimal loss in training accuracy.

\paragraph{Comparison with Zero-Offload} Compared to Zero-Offload, \SysName achieves faster convergence on Alpaca. As shown in Fig.~\ref{fig:gpt2} and~\ref{fig:llama3b}, \SysName uses around 62.5\% and 33.1\% less time when converging to the same accuracy. E.g., when training on the Laptop GPU, \SysName achieves the evaluation perplexity of 1.82 after 2 hours of training, while reaching the same perplexity takes 4.5 hours with Zero-Offload. Also, \SysName converges to the perplexity of 1.63 after 12 hours, which is achieved by Zero-Offload after 20 hours. Moreover, as shown in Fig.~\ref{fig:deepseek-coder} and Tab.~\ref{tab:coding}, within the 120 hour training budget, \SysName trains 1.97x more epochs than Zero-Offload, resulting in lower training losses.
Similarly, for the Deepseek-Coder-6.7B model, \SysName completes the fine-tuning for one epoch 2x faster than Zero-Offload while achieving close accuracy. When trained for 15 hours, \SysName outperforms Zero-Offload on average accuracy by $2.4\%$.

\begin{figure}[t]
     \centering
     \begin{subfigure}[b]{0.24\textwidth}
         \includegraphics[width=\textwidth]{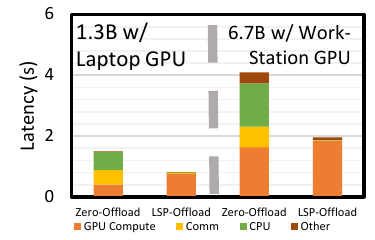}
         \caption{Breakdown for training 1 iteration on the coding task.}
         \label{fig:deepseek-coder-breakdown}
     \end{subfigure}
     \begin{subfigure}[b]{0.22\textwidth}
         \centering
         \includegraphics[width=\textwidth]{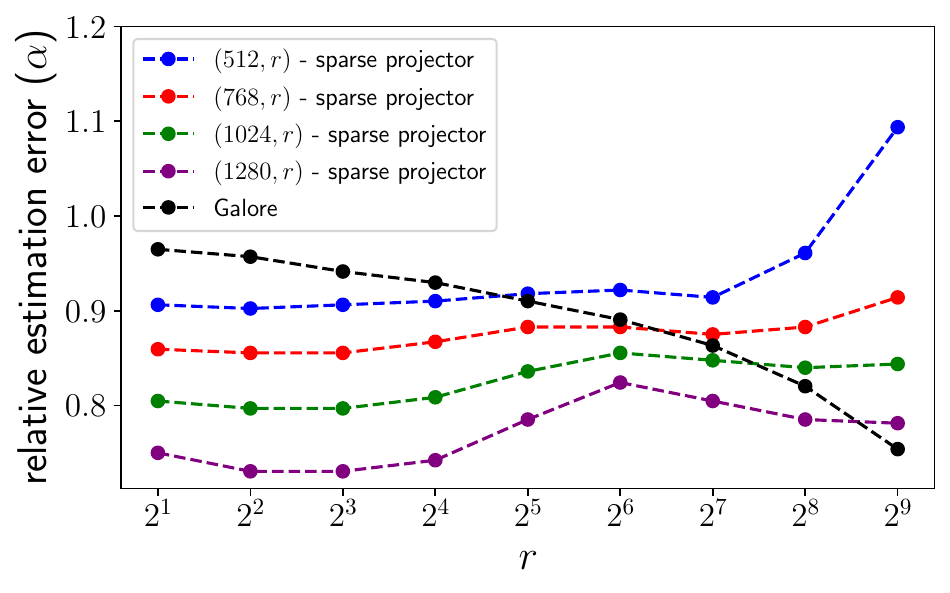}
         \caption{Empirical estimation bias w/ Deepseek-1.3B Model on the validation dataset.}
         \label{fig:hyperparameter_testloss}
     \end{subfigure}
     \vspace{-0.10in}
    \caption{Analysis on Coding task.}
    \vspace{-0.25in}
\end{figure}

\paragraph{Comparison with PEFT approaches} Furthermore, \SysName achieves 30\% lower evaluation perplexity than LoRA in Alpaca (Fig.~\ref{fig:gpt2}),
and outperforms GaLore in all coding tasks with 27.8\% higher average accuracy on the Humaneval ~\cite{chen2021evaluatinglargelanguagemodels,multiple10103177} dataset (Tab.~\ref{tab:coding}), even if GaLore trains 60\% more epochs than \SysName. As shown in Fig.~\ref{fig:hyperparameter_testloss}, this is because \SysName's sparse projector has lower estimation bias as compared to Galore's.

\begin{figure}[t]
\centerline{\includegraphics[width=\linewidth]{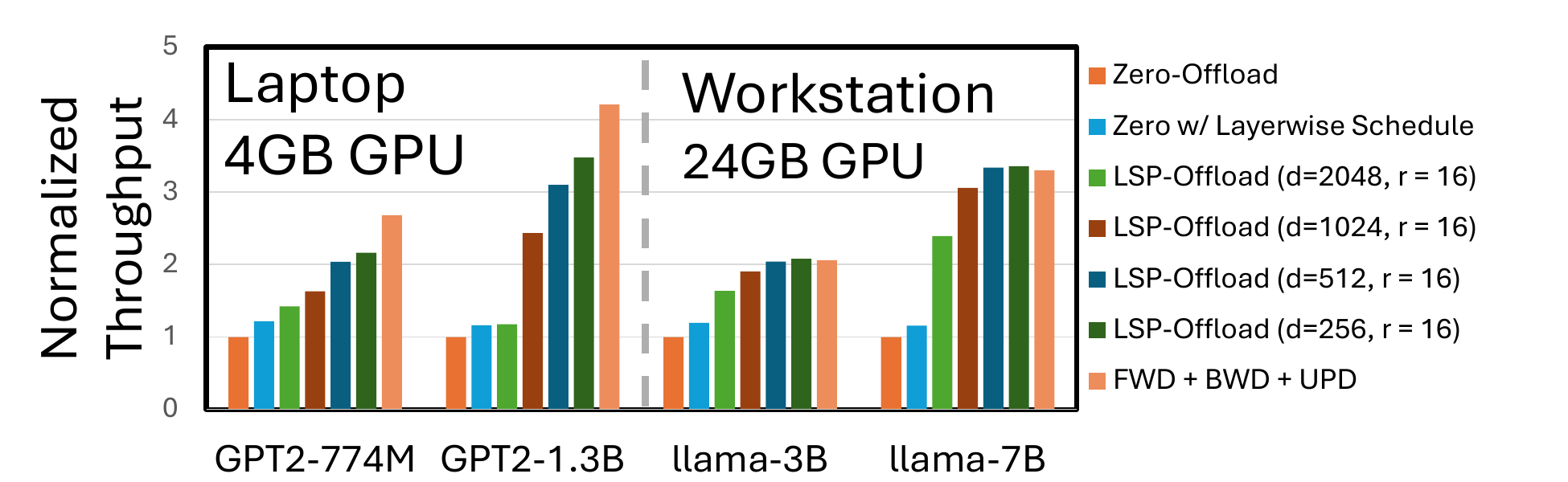}}
         \vspace{-0.05in}
         \caption{Training throughput comparison.}
         \label{fig:throughput}
         \vspace{-0.15in}
\end{figure}

\paragraph{Ablation Study.} Fig.~\ref{fig:throughput} shows an ablation study on training throughput with different techniques. Training throughput is measured by the number of training iterations executed in unit time. First, by adding layer-wise scheduling (blue columns), we improve Zero-Offload (leftmost column) throughput by 18\%. After that, we apply \SysName with different configurations. Compared to a native training setup (rightmost column) where only FWD, BWD, and UPD operations are performed on the GPU without CPU computation or communication, \SysName incurs an average slowdown of just 10.6\%, 16.7\% for subspace sizes of 256, 512 respectively. 

\paragraph{Training time breakdown.} Fig.~\ref{fig:deepseek-coder-breakdown} shows the time breakdown of \SysName for training a single iteration. Compared to Zero-Offload, \SysName cuts 50\% the per-iteration latency by reducing the wall-clock time of CPU compute and communication. Because of the layer-wise parallel schedule, the communication and compute on both CPU and GPU are fully paralleled, resulting in minimal non-overlapped overhead for communication and CPU compute.

\paragraph{Hyperparameter.} 
We measured the estimation bias across different configurations on the Deepseek Coding task. As shown in Fig.~\ref{fig:hyperparameter_testloss}, increasing $d$ consistently reduces estimation bias, leading to faster and higher quality convergence. Therefore, it is advisable to set $d$ as large as possible, provided that communication is not becoming a bottleneck. Empirically, we found that setting $d$ to half the size of the matrix is an effective balance between performance and overhead. For the rank $r$, we found that smaller $r$, such as 4 or 8, tends to result in better generalization. This is particularly favorable because it introduces minimal storage overhead while maintaining the training performance. 


\section{Limitation}
\SysName introduces a few hyperparameters that matter for the best performance, including the selection of the $(d,r)$-sparse projector, the frequency of subspace updates, the threshold for these updates, and others. 
Additionally, while compression and decompression may introduce compute overhead on the GPU, this can be effectively mitigated by implementing specialized \textit{sparse}-matrix multiplication kernels, which we also intend to address in future work.

\section{Conclusion}
In this paper, we address the bottlenecks of communication and CPU compute in current offloading frameworks. Inspired by PEFT methods, we developed \SysName to enable near-native speed fine-tuning by constraining parameter updates to a subspace. Using a sparse projector and minimizing empirical bias, \SysName optimizes in larger spaces than GaLore and LoRA with the same GPU memory. On the GLUE dataset, \SysName achieves convergence at the same rate as native training. Compared to Zero-Offload, it reduces fine-tuning time by 33.1\%–62.5\% on instruction-tuning tasks while maintaining accuracy. Additionally, it improves accuracy by 27.8\%–30\% on Alpaca and Humaneval datasets over GaLore and LoRA.

\section{Acknowledgment}
This work is supported by grants from the National
Science Foundation (NSF CNS-2211882), and by the member companies of the
Wasm Research Center and PDL consortium. We thank Aashiq Muhamed, David Woodruff, and Dave Anderson for the discussion on this work, as well as anonymous reviewers, for providing valuable feedback.
\section{Appendix}


\begin{figure*}
    \centering
    \includegraphics[width=\textwidth]{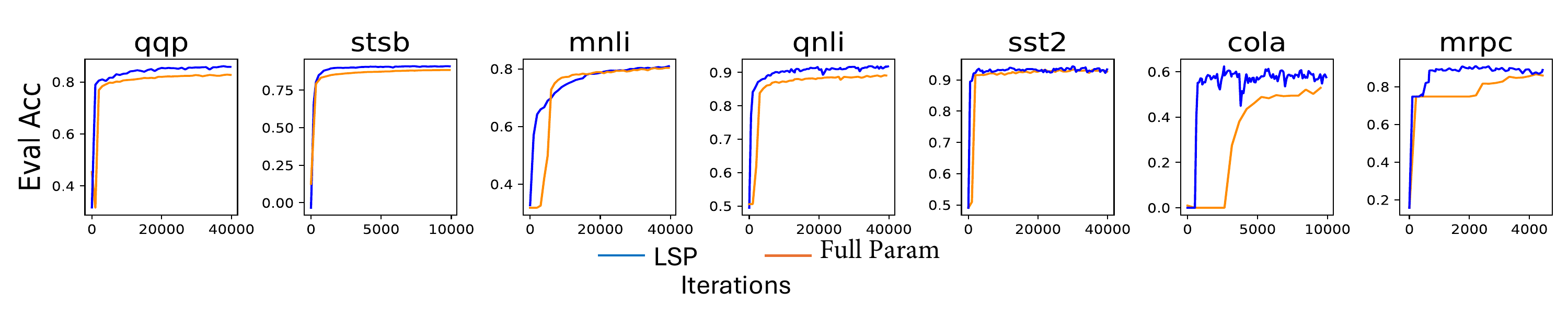}
    \vspace{-0.2in}
    \caption{Convergence Validation of LSP by finetuning pre-trained RoBertA-base model on GLUE.}
    \label{fig:glue}
\end{figure*}

\subsection{Zero's Schedule}

\begin{algorithm}
\caption{Zero-Offload's Pseudo-code}\label{alg:zero}
\begin{algorithmic}[1]
\STATE \textbf{Input:} $\GPUmodel$: GPU model, ${\cal D}$: Dataset, $S$: Optimizer state on CPU, $W$: Weights
\FOR{$t \gets 1$ to $\tau$}
    \STATE Sample $x \sim {\cal D}$
    \STATE $l \gets \GPUmodel.forward(x)$\COMMENT{FWD on GPU}
    \STATE $\nabla_Wl \gets SendToCPU(\GPUmodel.backward(l))$\COMMENT{Paralleled BWD on GPU and Gradient Offload}
    \STATE $\Delta W \gets SendToGPU(Update(\nabla_Wl, S))$\COMMENT{Paralleled Update on CPU and Delta Upload}
    \STATE $W\gets W+ \eta_t\cdot\Delta W$\COMMENT{on GPU, learning rate $\eta_t$}
\ENDFOR
\end{algorithmic}
\end{algorithm}

\subsection{Properties of $(d,r)$-Sparse Projectors}
\begin{property}[commutativity]
For a distribution $\mathcal{D}$ on $R^{m\times n}$ matrices, P, Q the sparse projector,
\begin{align}
    E_{\Sigma \sim \mathcal{D}}[P^T(\Sigma)Q] &= P^T(E[\Sigma])Q\\
    E_{\Sigma \sim \mathcal{D}}[\textbf{b}^{P,Q}(\Sigma)] &= \textbf{b}^{P,Q}(E[\Sigma]) \label{eqn:commutativebias}
\end{align}
\end{property}

\subsection{Proof of Theorem~\ref{the:convergence}}

Our proof adapts the analysis in~\cite{stich2020analysis}.

Before proving Theorem~\ref{the:convergence}, we list the lemmas used in the proof.

\begin{lemma}[Matrix Chernoff]\label{lemma:matrix chernoff}
    Let $M_1, ..., M_t$ be independent matrix valued random variables such that $M_i \in \mathbb{R}^{d_1\times d_2}$ and $\mathbb{E}[M_i] = 0$. If $\|M_i\|_2 \le \gamma$ holds almost surely for all $i\in \{1,...,t\}$, then for every $\epsilon > 0, 0 < \delta < 1$, when $t > \frac{8\gamma^2}{3\epsilon^2}\log(\frac{d_1+d_2}{\delta})$,
    \[Pr(\|\frac{1}{t}\Sigma_i M_i \|_2>\epsilon) \le \delta\]
\end{lemma}

\begin{lemma}\label{lemma:bias}
    For sparse projector $P,Q$, under Assumption~\ref{asp:boundedBias}, we can bound the bias by the empirical bias on a random sub-sampled dataset $S$ of size $\mathcal{O}(\frac{8\gamma^2}{3\epsilon^2}\log{\frac{m+n}{\delta}})$  with probability at least $1 - \delta$, 
    \[\|\textbf{b}^{P,Q}(\nabla f(W))\|_2 \le \|\textbf{b}^{P,Q}(\nabla f_S(W)) \|_2 + \epsilon, \]
    where $f_S (W) := \Sigma_{x\sim S} f_x(W) / |S|.$
\end{lemma} 

\begin{proof}
    For data $x \in S$, let $M_x = \textbf{b}^{P,Q}(\nabla f_x(W)) - \textbf{b}^{P,Q}(\nabla f(W))$. By the commutativity of the bias (eqn.~\ref{eqn:commutativebias}), $\mathbb{E}[M_x] = \textbf{0}$. Under Assumption~\ref{asp:boundedBias}, $\|M_x\| < \gamma$. Also, 
    \begin{align*}
    \frac{1}{|S|}\Sigma_{x\in S} M_x &= \frac{1}{|S|}\Sigma_{x\in S} (\textbf{b}^{P,Q}(\nabla f_x(W)) - \textbf{b}^{P,Q}(\nabla f(W))) \\ 
                                     &= \frac{1}{|S|}\Sigma_{x\in S} (\textbf{b}^{P,Q}(\nabla f_x(W)) ) - \textbf{b}^{P,Q}(\nabla f(W)) \\ 
                                     &= \textbf{b}^{P,Q}(\frac{1}{|S|}\Sigma_{x\in S} (\nabla f_x(W))) - \textbf{b}^{P,Q}(\nabla f(W)) \\ 
                                     &= \textbf{b}^{P,Q}(\nabla f_S(W)) - \textbf{b}^{P,Q}(\nabla f(W)).
    \end{align*}
    By Matrix Chernoff (lemma~\ref{lemma:matrix chernoff}), we have that for $|S| > \frac{8\gamma^2}{3\epsilon^2}\log{\frac{m+n}{\delta}}$,
    \[Pr(\|\textbf{b}^{P,Q}(\nabla f_S(W)) - \textbf{b}^{P,Q}(\nabla f(W)) \|_2>\epsilon) \le \delta.\]
    Therefore, with probability $1 - \delta$, 
     \begin{align*}
     \|\textbf{b}(\nabla f(W))\|_2  &\le \|\textbf{b}^{P,Q}(\nabla f_S(W))\|_2 + \\ &\|\textbf{b}^{P,Q}(\nabla f_S(W)) - \textbf{b}^{P,Q}(\nabla f(W)) \|_2  \\
     &\le \|\textbf{b}^{P,Q}(\nabla f_S(W))\|_2 + \epsilon
     \end{align*}
\end{proof}

\begin{theorem}\cite{stich2020analysis}\label{thm:bias convergence}
    For any $\epsilon > 0, 0 < \delta < 1$, suppose f is an L-smooth function\footnote{A function f: $\mathbb{R}^d\to \mathbb{R}$ is an L-smooth function if it is differentiable and there exists a constant $L > 0$ such that $f(\textbf{y}) \le f(\textbf{x}) + \langle\nabla f(\textbf{x}), \textbf{y} - \textbf{x}\rangle + \frac{L}{2}\|\textbf{y}-\textbf{x}\|^2$.}, and for any weight matrices $W$, $\|\textbf{b}^{P,Q}(\nabla f(W))\| \le m\nabla f(W) + \psi$, where $0 < m < 1, \psi > 0$, and stepsize $\eta = \frac{1}{L}$. Denote $F:= \mathbb{E}[f(W_0)] - f^*$, Then with probability $1 - \delta$, 
    \[\tau=\mathcal{O}(\frac{1}{\epsilon})\cdot \frac{LF}{(1-m)}\]
    iterations are sufficient to obtain $\min_{t\in[T]}\mathbb{E}\|\nabla f(W_t)\|^2 = \mathcal{O}(\epsilon + \frac{\psi}{1 - m}).$
\end{theorem}

Now, we prove Theorem~\ref{the:convergence}.
\begin{proof}
We analyze with some $\delta_0 > 0$ and $\beta > 0$.
From lemma~\ref{lemma:bias}, under the Assumption~\ref{asp:boundedBias}, we know that for $|S| > \frac{8\gamma^2}{3\beta^2}\log{\frac{m+n}{\delta_0}}$, with probability $1-\delta_0$,  
\[\|\textbf{b}^{P,Q}(\nabla f(W))\|_2 \le \|\textbf{b}^{P,Q}(\nabla f_S(W))\|_2 + \beta. \]

Also, because $\mathbb{E}[\nabla f_S(W)] = \nabla f(W)$ (lemma~\ref{lemma:matrix chernoff}), under the Assumption~\ref{asp:boundedBias}, we have for $|S| > \frac{8\gamma^2}{3\beta^2}\log(\frac{d_1+d_2}{\delta_0})$, with probability $1-\delta_0$, 
\[\|\nabla f_S(W) - \nabla f(W)\|_2 \le \beta \]

We bound the bias for every parameter update step,
\begin{align*}
    \|\textbf{b}^{P,Q}(\nabla f(W))\|_F &\\
    \text{by Assumption~\ref{asp:sparseBias}} &\le c\|\textbf{b}^{P,Q}(\nabla f(W))\|_2 \\ 
            \text{with prob}~1-\delta_0&\le c\|\textbf{b}^{P,Q}(\nabla f_S(W))\|_2 + c \beta \\ 
            \text{by Assumption~\ref{asp:effOfSubspace}}            &\le c\alpha\|\nabla f_S(W)\|_2 + c\beta \\
            \text{with prob}~1-\delta_0&\le c\alpha\|\nabla f(W)\|_2 + c\beta + c\alpha \beta \\ 
                        &\le c\alpha\|\nabla f(W)\|_F + c\beta + c\alpha \beta \\ 
\end{align*}
Thus, with probability $1 - 2\delta_0$, 
\[\|\textbf{b}^{P,Q}(\nabla f(W))\|_F^2 \le 2c^2\alpha^2\|\nabla f(W)\|_F^2 + 2c^2\beta^2(1+\alpha)^2.\]
By plugging this into Theorem~\ref{thm:bias convergence} for all steps from 1 to $\tau$, we have that for $|S| > \frac{8\gamma^2}{3\beta^2}\log(\frac{d_1+d_2}{\delta_0})$, with probability $1 - 2\tau\delta_0$, 
\[\tau=\mathcal{O}(\frac{1}{\epsilon})\cdot \frac{LF}{(1-2c^2\alpha^2)}\]
iterations are sufficient to obtain $\min_{t\in[\tau]}\mathbb{E}\|\nabla f(W_t)\|^2 = \mathcal{O}(\epsilon + \frac{2c^2\beta^2(1+\alpha)^2}{1 - 2c^2\alpha^2})$. Setting $\delta_0 = \frac{\delta}{2\tau}$ concludes the proof. 
\end{proof}

\subsection{Layer-wise Scheduling}

\begin{table*}
\centering
\caption{Configurations and timings for training/fine-tuning the GPT2-1.3B Model (using fp16) on commodity laptop hardware---the Nvidia A1000 GPU (4GB) and Intel Core-i7 12800H CPU (32GB).}
\label{tab:gpt2-1b}
\vspace{-0.05in}
\begin{tabular}{llllll}
\hline
\rowcolor{Gray}
Parameters & Optimizer State & Activations & CPU-GPU Bandwidth & \#Layers & GPU Memory \\ \hline
2.6GB        & 7.8GB             & 0.5GB         & 10--15GB/s            & 40       & 4GB\\
\hline\hline
\rowcolor{Gray}
FWD on CPU & BWD on CPU & UPD on CPU & FWD on GPU & BWD on GPU & UPD on GPU\\\hline
0.16s/layer & 0.27s/layer & 0.08s/layer & 4.5ms/layer & 8.7ms/layer & 7.9ms/layer \\
\hline
\end{tabular}
\end{table*}

\begin{algorithm}[H]
\caption{Layer-wise Scheduling}\label{alg:LayerwiseSch}

\begin{algorithmic}[1]
\STATE {\textbf{Hyperparameter: }} $TransitionLayer:$ prevlayer to change the schedule mode from FirstComeFirstServe to LastComeFirstServe. Others are same as Alg.~\ref{alg:SGESchedule}.
\FOR{$t \gets 0$ to $\tau - 1$}
    \STATE Sample $(x_0, y) \sim \mathcal{D}$
    \FOR{$l$ in $layers$}
        \STATE Wait for event $e_l$ \COMMENT{prevforward pass happens after the parameter gets updated}
        \STATE $x_l \gets \text{forward}(x_{l-1}, l, W_l)$
    \ENDFOR
    \STATE $grad = \text{loss}(x_l, y)$
    \STATE $mode \gets \text{FCFS}$
    \FOR{$l$ in $\text{reversed}(layers)$}
        \IF{$l$ is $TransitionLayer$} 
            \STATE $mode \gets \text{LCFS}$
        \ENDIF 
        \STATE $grad, \nabla_{W_l} \gets \text{backward}(grad, x_l, l, W_l)$
        \STATE $\hat{\nabla}_{W_l} \gets  P^T_l\nabla_{W_l}Q_l$
        \STATE $\text{AsyncMemcpy}(mode, \hat{\nabla}_{W_l}, GPU2CPU)$ 
        \STATE $\text{AsyncExecOnCPU}( mode, \Delta_{W_l} \gets Update(\hat{\nabla}_{W_l}))$
        \STATE $\text{AsyncMemcpy}(mode, \Delta_{W_l}, CPU2GPU)$
        \STATE $\text{AsyncExecOnGPU}(mode, W_l \gets W_l - \eta_t P_l\Delta_{W_l} Q_l^T, CPU2GPU)$
    \ENDFOR
\ENDFOR
\end{algorithmic}
\end{algorithm}

\paragraph{Avoiding blocking.} To avoid the deeper layer's workload from blocking the shallower layer's computation that executes earlier in the next iteration, we use a heuristic to switch between two schedule mode: $FirstComeFirstServe$ (FCFS) and $LastComeFirstServe$ (LCFS). When the backward pass begins, FCFS is used first for parallelizing GPU compute and offloading. As the backward pass proceeds, we change the Schedule to LCFS which helps shallower layers get ready for the next pass. We set the switch point to be $TransitionLayer = \#Layer - \frac{T_{BWD} - (T^{layer}_{Offload}+ T^{layer}_{Upload}+ T^{layer}_{UPD})}{\max\{T^{layer}_{Offload}, T^{layer}_{Upload}, T^{layer}_{UPD}\}}$, which is the deepest layer that may block the computation of the first layer.

\subsection{Implementation}
We prototyped \SysName as a Python library built on top of Pytorch. \SysName can automatically detect each matrix multiplication module and replace it with the offloaded version. To achieve best performance, we implemented the fused Adam kernel in Zero-Offload to accelerate the parameter update on the CPU. Also, we used a pinned memory buffer on the CPU to enable fast communication, and used CUDA streams for paralleled communication and computation. Moreover, gradient checkpoint is enabled to reduce the activation memory. 



\subsection{Experiment Configurations and Further Results}\label{apx:hyper param}

As noted in the Evaluation section, our \textbf{laptop GPU} setup is an Intel Core-i7 12800H CPU (32GB) laptop with an Nvidia A1000 Laptop GPU (4GB), and our \textbf{workstation GPU} setup is an AMD Ryzen Threadripper 3970X CPU (252GB) with an Nvidia RTX 4090 GPU (24 GB).

For all experiments, the random seed is set to a fixed value shown in the code.

\subsubsection{GLUE Experiment}
For the GLUE experiment, we use a batch size of 16 and a learning rate of 5e-5 for \SysName and Full Parameter fine-tuning. For \SysName, we update the subspace at the beginning of each epoch or every 1000 iterations ($CheckFreq=1000$) and set $\alpha$, the threshold for updating projectors, to 0.3. All experiments are limited to train for 1 hour. 

As shown in Fig.~\ref{fig:glue}, for all cases in GLUE, Alg.~\ref{alg:SGESchedule} is able to converge at a comparable rate \textit{per iteration} as full parameter fine-tuning, despite its use of lossy compression (learned sparse projectors). 
Since full parameter fine-tuning suffers from significantly slower iteration times than LSP, the convergence rate \textit{per hour} is slower than LSP.
As Tab.~\ref{tab:glue} showed, this results in LSP achieving 0.855 average accuracy compared to Full Parameter's 0.836 after 1 hour.

\subsubsection{Instruction Fine-tuning on Alpaca}
For the instruction fine-tuning experiments on the Alpaca data set, we use a batch size of 4 for the GPT2-774M model and 16 for the Llama-3B model, which are the largest without exceeding the laptop GPU and workstation GPU memory, respectively. We set the learning rate to be the best among \{1e-4, 1e-5, 1e-6\}, which is 1e-4 for \SysName and 1e-5 for both LoRA and Zero-Offload. For \SysName, $CheckFreq=1000$ and $\alpha=0.5$. 

\subsubsection{Instruction Fine-tuning with DeepSeek-Coder}
For the code instruction fine-tuning experiments, we use the gradient accumulate technique to simulate large batch sizes by averaging gradients from multiple small batches before updating the weights.

For the DeepSeek-Coder-1.3B model, we chose a simulated batch size of 128, max sequence length of 1024, the AdamW optimizer, max epoch number of 5, and the Cosine learning rate scheduler. The rank of LoRa is selected as the maximum value that the GPU memory can accommodate. The LoRa alpha ($\alpha$) for LoRa (Rank=8) is 32. Because GaLore does not perform well with ranks that are small enough to fit into the GPU memory, we chose GaLore's rank to be 256, which uses 7.9GB memory. The alpha ($\alpha$) for GaLore (Rank=256) is the default value of 0.25 in their library. 
We tried different learning rate from 1e-5 to 2e-4 and the learning rate is set to be the optimal value across multiple experiments.
We found that the learning rate of 1e-4 performed well across different settings and used it in our final experiments. 
The evaluation accuracy of fine-tuning DeepSeek-Coder-1.3B on the Humaneval dataset shown in Tab.~\ref{tab:coding}(top) are the scores corresponding to the checkpoint after approximately 120 hours of training on the laptop GPU.

For the DeepSeek-Coder-6.7B model, we chose a simulated batch size of 64, max sequence length of 1024, the AdamW optimizer, max epoch number of 1, and the Cosine learning rate scheduler with minimal learning rate. We found that the learning rate of 1e-4 and minimal learning rate of 5e-5 performs well across different learning rate settings.
The evaluation accuracy of fine-tuning DeepSeek-Coder-6.7B shown in Tab.~\ref{tab:coding}(bottom) are the scores corresponding to the checkpoint after approximately 15 hours of training on the workstation GPU.
The table also shows the accuracy of Zero-Offload after 30 hours of training.

For the DeepSeek-Coder experiments, due to the extended training time required with offloading, we simulate the training process by:
\begin{enumerate}
    \item Training on a GPU with sufficient memory: This allows us to obtain the training performance (e.g., training loss, evaluation score, etc.) as a function of training steps.
    \item Profiling the average time per training step with offloading under maximum supported token batch size under the memory limit.
    \item Combining the results: We merge the performance data from step 1 and scale the token batch size in step 2 to match the accumulated token batch size per update step in step 1, to map training performance against training time.
\end{enumerate}
This approach enables us to simulate the training performance over time as if offloading were being used, without needing to actually train with offloading. Specifically, for both \SysName and Zero-Offload, on the DeepSeek-Coder-1.3B (6.7B) experiment, we chose the token batch size as $384=1\times 384$ ($4096=4\times 1024$) respectively. The per-iteration profile result is shown in Fig.~\ref{fig:deepseek-coder-breakdown}. 


\subsection{\SysName's Hyperparameters}\label{apx: hyper-parameter}

\begin{figure}
    \centering
    \begin{subfigure}{0.23\textwidth}
    \includegraphics[width=\linewidth]{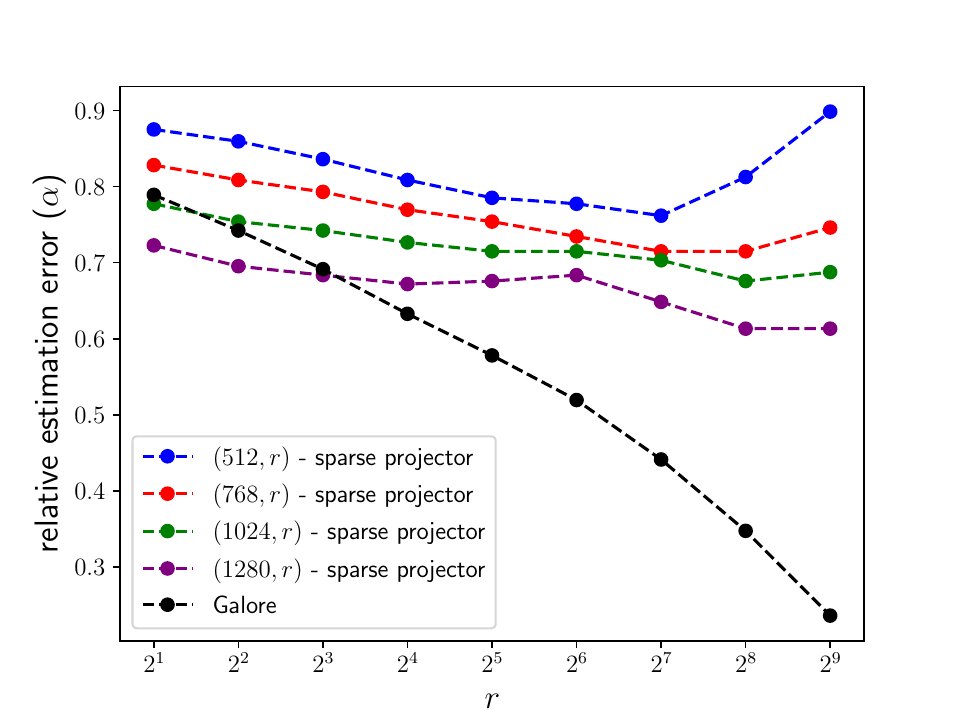}
    \caption{Estimation Error on Calibration Dataset}
    \label{fig:hyperparameter_train_loss}
    \end{subfigure}
    \begin{subfigure}{0.23\textwidth}
    \includegraphics[width=\linewidth]{figures/hyperparameter_test_loss.pdf}
    \caption{Estimation Error on Validation Dataset}
    \label{fig:hyperparameter_test_loss}
    \end{subfigure}
    \caption{Benchmark on Estimation Error for Deepseek Coding-1.3B fine-tuning task}
    \label{fig:hyperparameter_apx}
\end{figure}

\SysName introduces some new hyperparameters: including the number of non-zero values $r$ in each row of the $(d,\ r)$-sparse projector, the size $d$ of the subspace, and the frequency $CheckFreq$ and $\alpha$ threshold of the projector update. Among them, $d$ and $r$ have a great influence on the effect of the projector.

We tested the $estimated\ bias$ with our learned sparse projectors and with the orthogonal projectors used in GaLore on the DeepSeek-Coder-1.3B fine-tuning task. The orthogonal projectors are the spectrum of $\nabla_W$ calculated via Singular Value Decomposition (SVD):
\begin{equation}
\begin{aligned}
& \nabla_W = USV^T \approx \sum^r_{i=1} s_i u_i v_i^T \\
& P = [u_1,u_2,\dots,u_r],\quad Q=[v_1,v_2,\dots,v_r]^T
\end{aligned}
\end{equation}
where $r$ is also called the rank of the orthogonal projectors. We will use GaLore($r$) to indicate GaLore run with rank=$r$. 
Recall from Tab.~\ref{tab:method comparisons} that for the same $r$, GaLore($r$) uses more GPU memory than \SysName with $(d,r)$-sparse projectors.
The results are shown in Fig.~\ref{fig:hyperparameter_apx}.

Although Fig.~\ref{fig:hyperparameter_train_loss} shows that GaLore($r$) has a lower \textit{training} error than $(d, r)$-sparse projectors for the same $r$ when $r \geq 16$,
$(d, r)$-sparse projectors generalize much better: their \textit{test} errors are much lower than GaLore's.
For example, Fig.~\ref{fig:hyperparameter_test_loss} shows that $(1280, r)$-sparse projectors achieve lower test error than GaLore($r$) for the same $r$ for all $r \le 256$. 
We attribute this to $(d, r)$-sparse projectors' decoupling of the subspace size $d$ and the non-zero values $r$ (compared to GaLore's sole use of a rank $r$), such that we can optimize in a large subspace with minimal extra GPU memory. 

Secondly, we found the performance of the learned $(d, r)$-sparse projectors does not necessarily improve as $r$ increases. On the contrary, selecting a relatively small $r$, such as 4 or 8, tends to result in better generalization for this fine-tuning task (while also reducing GPU memory usage and shortening iteration times). 

At the same time, the effectiveness of the projector improves as the subspace size $d$ increases. Therefore, it is advisable to select the largest possible subspace size that does not cause an unacceptable decrease in training performance or exceed the GPU memory capacity.

\end{document}